# Utility Optimal Scheduling in Degree-Limited Satellite Networks

LI Hu, LIU Yuan'an, YUAN Dongming, HU Hefei, DUAN Sirui

*School of Electronic Engineering, Beijing University of Posts and Telecommunications (BUPT), Beijing 100876, China*

## Abstract

In this paper, we consider the problem of flow control together with power allocation to antennas on satellite with arbitrary link states, so as to maximize the utility function while stabilizing the network. Inspired by Lyapunov optimization method, a Degree-Limited Scheduling Algorithm (DLSA) is proposed with a control parameter V, which requires no stochastic knowledge of link state. Discussion about implementation is carried out about the complexity of DLSA and several approximation methods to reduce complexity. Analyze shows DLSA stabilizes the network and the gap between utility function under DLSA and the optimal value is arbitrarily close to zero on the order of O(1/V). Simulation results verify DLSA on a simple network.

## 1. Introduction

Satellite networks can play an important role in providing ubiquitous communication for high data rate applications in the future. However resources in satellite networks are limited but traffic and delay demands are increasing rapidly. [1] Therefore, effective utilization of system resources is the key technical issue for satellite networks. In this environment, it is important to decrease the delay in network while improving the total network capacity.

In recent years, to achieve high performance in satellite networks, numerous algorithms have been proposed by various methods. [2] proposes an algorithm aiming at reducing power while preserving certain network capacity. [3] addresses an optimal bandwidth allocation algorithm for networks with large propagation delay which can be applied to satellite systems. [4] describes a power allocation algorithm for downlinks of a multibeam satellite which is able to make a compromise between total capacity maximization and uses' fairness. [5] proposes an algorithm jointly optimizing resource allocation and congestion control for satellite systems with a multiple beam antennas and phased array antennas. [6] addresses a power allocation algorithm aiming at that the number of subscribers not receiving the desired quality of service is minimized. [7] describes a resource allocation scheme for multi-beam satellite networks that can dynamically offer maximum communication capacity without compromising quality. However, most of the existing works focus on downlink scenarios or single-hop networks, and often require sufficient statistical knowledge, without taking the limitation of in-degree and out-degree of a satellite into consideration.

In this paper, we proposed a joint schedule algorithm for a specific type of satellite networks where inter-satellite links are established by full-duplex directional microwave antennas or laser transmitters. This type of systems is different from territorial communication systems, satellite networks without inter-satellite links, and satellite networks with inter-satellite links but omni-directional antennas, hence there are multiple special factors should be noticed when designing algorithms for such a system. Firstly, the state of link between satellites is time-varing because interference and attenuation are changing because of movement of satellites. Secondly, power budget is relatively tight because energy of satellites is harvested by using solar panels.

Thirdly, antennas in one satellite are relatively few, leading to the fact that a satellite may not able to communicate with all visible satellites, which means the in-degree and out-degree of a satellite are bounded. Lastly, two satellites can communicate only when they point their antennas to each other simultaneously. [8]

Our work differs from others (e.g. [2]-[7]) in the way that we consider all the factors listed above together, especially the last two, which is rarely considered in other works. We tackle this problem using the Lyapunov optimization technique developed in [9] and [10]. The idea of this approach is to construct the algorithm based on a quadratic Lyapunov function, "pushing" the target queue levels towards zero to stabilize all queues of network while satisfying some timing-average constraints. Based on this idea, we construct the Degree-Limited Scheduling Algorithm, which is a joint algorithm with data admission, power allocation, link selection and routing and requires no statistical knowledge of the channel qualities.

This paper is organized as follows: In Section 2 we state our network model and the objective problem. In Section 3, we present the DLSA algorithm. In Section 4, we analyze the performance of DLSA, deriving the exact bound of utility function. Simulation results are presented in Section 5.

## 2. The Network Model

### 2.1 Network & Utility Function

We consider a time-slotted multihop satellite network where timeslot $t$ corresponds to the time interval $[t, t+1)$. There are $N$ nodes (satellites) in the network, consisting the node set $\mathcal{N}$. $\mathcal{L} = \{(i,j) \mid i, j \in \mathcal{N}\}$ denotes the set of links between nodes. The network is modeled by a directed graph $\mathcal{G} = \{\mathcal{N}, \mathcal{L}\}$. For each node $n$, $\mathcal{N}_n^o$ denotes the set of nodes $\{b \mid b \in \mathcal{N}, (n,b) \in \mathcal{L}\}$, $\mathcal{N}_n^i$ denotes the set of nodes $\{a \mid a \in \mathcal{N}, (a,n) \in \mathcal{L}\}$. $d^o = \max_{n \in \mathbf{N}} |\mathcal{N}_n^o|$ is defined as the maximum out-degree of any node. $d^i = \max_{n \in \mathbf{N}} |\mathcal{N}_n^i|$ is defined as the maximum out-degree of any node.

There are $C$ types of packet in the network, where the destination of packets of type $c$ is node $c$. At timeslot $t$, the network decides the amount of packets of type $c$ admitted into node $n$, denoted as $R_n^c(t)$. We assume that $0 \le R_n^c(t) \le R_{\max}$ $\forall n, c, t$, with some finite $R_{\max}$. Utility function is defined as $U_{tot}(\mathbf{r}) = \sum_{n,c} U_n^c(\overline{r}^{nc})$, where $\overline{r}^{nc}$ is the time average rate of packets of type $c$ admitted into node $n$, defined as $\overline{r}^{nc} = \limsup_{t \to \infty} \frac{1}{t} \sum_{\tau=0}^{t-1} \mathbb{E}\{R_n^c(\tau)\}$. Each $U_n^c(\overline{r}^{nc})$ is assumed to be increasing and strictly concave in $\overline{r}^{nc}$.

### 2.2 Channel

Channel state matrix $\mathbf{S}_{N \times N}(t) = (s_{ij}(t))|_{s_{ij}(t) \in \mathcal{S}}$ denotes the states of links at timeslot $t$, where set $\mathcal{S}$ contains all possible link states. We assume that $\mathbf{S}(t)$ takes values in some finite set $\{s_1, s_2, \ldots, s_M\}$. We assume that $\mathbf{S}(t)$ is i.i.d. every time slot and use $\pi_{s_i}$ to denote $\Pr(\mathbf{S}(t) = s_i)$.

At every timeslot, the network decides which links should be connected. Link connection matrix is denoted as $\mathbf{\Gamma}_{N \times N}(t) = (\gamma_{ij}(t))|_{\gamma_{ij}(t) \in \{0,1\}}$, where "1" stands for "connected", and "0" stands for "disconnected". Note that $\gamma_{ii}(t) = 0$ $\forall i, j, t$. Note that effective $\mathbf{\Gamma}(t)$ is symmetric because two nodes $i$ and $j$ can communicate at $t$ only when they both select the link $(i, j)$, otherwise power will be wasted and the transmission rate of other link will decrease since the total power is limited. Then we have $0 \le \sum_{j \in \mathcal{N}_i^o} \gamma_{ij}(t) \le d^o, 0 \le \sum_{i \in \mathcal{N}_i^i} \gamma_{ij}(t) \le d^i$ $\forall i, j, t$, which means at any timeslot, one node is able to transmit packets to at most $d^o$ node(s) and receive packets from at most $d^i$ node(s), i.e. the out-degree and in-degree of any node are bounded by $d^o$ and $d^i$, respectively.

At every timeslot, if $\mathbf{S}(t) = s_i$, then the power allocation matrix $\mathbf{P}_{N \times N}(t) = (p_{ij}(t))$ must be chosen from some finite power allocation set $\mathcal{P}^{\mathbf{S}(t)}$, where $p_{ij}(t)$ is the power allocated to link $(i, j)$ at time $t$. We assume $0 \le p_{ij}(t) \le P_{max}$ $\forall i, j, t$, with some finite $P_{max}$.

We assume $0 \le \limsup_{t \to \infty} \frac{1}{t} \sum_{\tau=0}^{t-1} \sum_{j \in \mathcal{N}_i^o} p_{ij}(\tau) \le P_i^{av}$, $\forall i$, with some finite $P_i^{av}$. This means that $P_i^{av}$ limits the maximum output power of node $i$. Intuitively, there is no benefit for allocating non-zero power for the link not connected. Although this argument is indeed correct, we choose to keep the current notation and reach the same conclusion through the policy specification itself rather than mere intuition.

## 2.3 Transmission Rate

Given the Channel state matrix $\mathbf{S}(t)$, link connection matrix $\mathbf{\Gamma}(t)$ and the power allocation matrix $\mathbf{P}(t)$, the transmission rate over the link $(i, j)$ is given by the rate-power function $\mu_{ij}(t) = \mu_{ij}(\mathbf{S}(t), \mathbf{\Gamma}(t), \mathbf{P}(t))$.

For each $s_i$, we assume that the function $\mu_{ij}(t)$ satisfies the following properties:

*Property 1:* For any matrix $\mathbf{P}(t), \mathbf{P}'(t) \in \mathcal{P}^{\mathbf{S}(t)}$, where $\mathbf{P}'(t)$ is obtained by changing any single component $p_{ij}(t)$ in $\mathbf{P}(t)$ to zero, we have for some finite constant $\delta > 0$:

$$\mu_{ij}(\mathbf{S}(t), \mathbf{\Gamma}(t), \mathbf{P}(t)) \le \mu_{ij}(\mathbf{S}(t), \mathbf{\Gamma}(t), \mathbf{P}'(t)) + \delta p_{ij}(t) \qquad (1)$$

This property states that the rate obtained over a link $(i, j)$ is upper bounded by some linear function of the power allocated to it.

*Property 2:* If $\mathbf{P}'(t)$ is obtained by setting the entry $p_{ib}(t)$ in $\mathbf{P}(t)$ to zero, then:

$$\mu_{aj}(\mathbf{S}(t), \mathbf{\Gamma}(t), \mathbf{P}(t)) \le \mu_{aj}(\mathbf{S}(t), \mathbf{\Gamma}(t), \mathbf{P}'(t)) \quad \forall (a, j) \ne (i, b) \qquad (2)$$

This property states that reducing the power over any link does not reduce the rate over any other links.

*Property 3*: If $\mathbf{P}'(t)$ is obtained by setting the entry $p_{ab}(t)$ in $\mathbf{P}(t)$ to zero, where $\gamma_{ab}(t) = 0$, then:

$$\mu_{ij}(\mathbf{S}(t), \mathbf{\Gamma}(t), \mathbf{P}(t)) = \mu_{ij}(\mathbf{S}(t), \mathbf{\Gamma}(t), \mathbf{P}'(t)) \qquad (3)$$

This property states that changing the value of $p_{ab}(t)$ where $(a, b)$ is disconnected, the resulting rate of any link will not be changed. We see that Property 1, 2 and 3 can usually be satisfied by most rate power functions. We also assume that there exists some finite constant $\mu_{max}$ such that $\mu_{ij}(t) \le \mu_{max}, \forall \mathbf{S}(t), \mathbf{\Gamma}(t), \mathbf{P}(t), \forall t$. In the following, we also use $\mu_{ij}^c(t)$ to denote the rate allocated to packets of type $c$ over link $(i, j)$ at timeslot $t$. It is obvious that $\sum_c \mu_{ij}^c(t) \le \mu_{ij}(t)$ $\forall i, j, t$.

## 2.4 Queue

Let $\mathbf{Q}(t) = (Q_n^c(t))$ be the data queue backlog vector in the network, where $Q_n^c(t)$ is the amount of packets of type $c$ queued at node $n$. We assume the following queuing dynamics:

$$Q_n^c(t+1) \le [Q_n^c(t) - \sum_{b \in \mathcal{N}_i^o} \mu_{nb}^c(t)]^+ + \sum_{a \in \mathcal{N}_i^i} \mu_{an}^c(t) + R_n^c(t) \qquad (4)$$

with $Q_n^c(t) = 0$ $\forall n, c$, $Q_c^c(t) = 0$ $\forall t$, and $[x]^+ = \max[x, 0]$. The inequality in (4) is due to the

fact that some nodes may not have enough commodity $c$ packets to fill the allocated rates.

In order to use the virtual queue technique introduce by [11], we construct the virtual queue $\mathbf{Z}(t) = (Z_n(t))$ to ensure the average output power is not greater than $P_{tot}$. We assume the following virtual queuing dynamics:

$$Z_n(t+1) = [Z_n(t) - P_{tot}]^+ + \sum_{b \in \mathcal{N}_n^o} p_{nb}\gamma_{nb} \qquad (5)$$

Using the definition in [12], we say that a discrete time process $X(t)$ is mean rate stable if:

$$\limsup_{t \to \infty} \frac{\mathbb{E}\{X(t)\}}{t} = 0 \qquad (6)$$

A network is stable i.f.f. all its data queues are stable.

## 2.5 Utility Maximization with Limited Degree

Under the network model described above, we can state our problem as follows.

$$\max \quad U_{tot}(\bar{\mathbf{r}}) \qquad (7)$$

$$\text{s.t.} \quad 0 \leq \sum_{j \in \mathcal{N}_i^o} \gamma_{ij}(t) \leq d^o, \ 0 \leq \sum_{i \in \mathcal{N}_i^i} \gamma_{ij}(t) \leq d^i, \ \gamma_{ij}(t) \in \{0,1\}$$

$$0 \leq \sum_{b \in \mathcal{N}_i^o} p_{nb}(t)\gamma_{nb}(t) \leq P_{tot}, \ 0 \leq p_{ij}(t) \leq P_{max}$$

All queues $Q_n^c(t)$ are stable.

This problem is not easy to solve. However, by using technique of virtual queues [11], the problem above can be transformed into the one below, which is solved by the DLSA proposed below based on Lyapunov optimization method.

$$\max \quad U_{tot}(\bar{\mathbf{r}}) \qquad (8)$$

$$\text{s.t.} \quad 0 \leq \sum_{j \in \mathcal{N}_i^o} \gamma_{ij}(t) \leq d^o, \ 0 \leq \sum_{i \in \mathcal{N}_i^i} \gamma_{ij}(t) \leq d^i, \ \gamma_{ij}(t) \in \{0,1\}, \ 0 \leq p_{ij}(t) \leq P_{max}$$

All queues $Q_n^c(t)$ and $Z_n(t)$ are stable.

## 3. Degree-Limited Scheduling Algorithm

### 3.1 DLSA

We now present the Degree-Limited Scheduling Algorithm (DLSA) with a control parameter $V$, which is inspired by [11] and [12]. The idea of the algorithm is to greedily minimize (18), which is the bound on Lyapunov drift of the network we study.

***Degree Limited Scheduling Algorithm (DLSA):***
At every timeslot $t$, observe $\mathbf{Q}(t)$, $\mathbf{Z}(t)$, $\mathbf{S}(t)$, and do the following:

**Data Admission**: At every time t, choose $R_n^c(t)$ to be the optimal solution of the following optimization problem:

$$\max \quad VU_n^c(r) - Q_n^c(t)r \qquad \text{s.t.} \quad 0 \leq r \leq R_{max} \qquad (9)$$

where $V$ can be any positive constant not less than 1.

**Power Allocation & Link Selection**: At every time t, choose $\Gamma(t)$ and $\mathbf{P}(t)$ to be the optimal solution of the following optimization problem:

$$\max \quad G(t) = \sum_{ij} (\mu_{ij}(p_{ij}(t), \gamma_{ij}(t))W_{ij} - Z_i(t)p_{ij}(t)\gamma_{ij}(t))$$

$$\text{s.t.} \quad 0 \leq \sum_{j \in \mathcal{N}_i^o} \gamma_{ij}(t) \leq d^o, \ 0 \leq \sum_{i \in \mathcal{N}_i^i} \gamma_{ij}(t) \leq d^i, \ \gamma_{ij}(t) \in \{0,1\}, \gamma_{ii}(t) = 0, \gamma_{ij}(t) = \gamma_{ji}(t) \qquad (10)$$

$$0 \leq p_{ij}(t) \leq P_{max}$$

where we define $W_{ij}^c(t) = [Q_i^c(t) - Q_j^c(t)]^+$ and $W_{ij}(t) = \max_c W_{ij}^c(t)$.

**Routing & Scheduling**: For every node $n$, find any $c_{ij}^*(t) = \arg\max_c W_{ij}^c(t)$. Then $\mu_{ij}^c(t)$ is given by:

$$\mu_{ij}^c(t) = \begin{cases} \mu_{ij}(t) & \text{if } c = c_{ij}^*(t) \text{ and } W_{ij} \geq 0 \\ 0 & \text{otherwise} \end{cases} \quad (11)$$

That is, allocate the full rate over the link $(i,j)$ to any commodity that achieves the maximum positive weight over the link. Use idle-fill if needed. If $W_{ij}(t) = 0$, we set $\mu_{ij}^c(t) = 0$ for all $c$ over link $(i,j)$.

**Queue Update**: Update $\mathbf{Q}(t)$ and $\mathbf{Z}(t)$ according to the dynamics (4) and (5), respectively.

## 3.2 Implementation of DLSA

Note that DLSA only requires the knowledge of the instant link state $\mathbf{S}(t)$, and the queue sizes $\mathbf{Q}(t)$ and $\mathbf{Z}(t)$. It does not even require any stochastic knowledge, which is very useful in practice when the stochastic knowledge is difficult to be obtained. Also note that if all the links do not interfere with each other, then DLSA can easily be implemented in a distributed manner, where each node only has to know about the queue sizes at its neighbor nodes and can decide on the power allocation locally.

The "Data Admission" part is a simple convex optimization problem, hence can be solved by methods proposed in [13]. Besides, if $U_n^c$ is chosen properly, this will reduce to "bang-bang" problem.

The "Power Allocation & Link Selection" part we can tell that DLSA's complexity is the same as the widely used max-weight algorithms, which in general requires centralized control and can be NP-hard [10], although the constraints here are all linear. And if $\boldsymbol{\mu}$ has a pleasant expression, this problem could be easy to solve. Besides, numerous heuristic algorithms can be used. Moreover, constant factor approximation solutions could be found in, e.g., [14] and Section 4.7 and 5.2.1 in [10]. Such approximation results can usually be found in a distributed manner in polynomial time.

## 4. Performance Analysis

To analyze the performance of DLSA, we first define the Lyapunov function as $L(t) = \frac{1}{2}\sum_{n,c}\left[Q_n^c(t)\right]^2 + \frac{1}{2}\sum_n\left[Z_n(t)\right]^2$.

Denote $\{\mathbf{Z}(t), \mathbf{Q}(t)\}$ as $\boldsymbol{\Theta}(t)$. Lyapunov drift function is defined as $\Delta(t) = \mathbb{E}\{L(t+1) - L(t) | \boldsymbol{\Theta}(t)\}$. Define $\Delta_V(t) = \Delta(t) - V\mathbb{E}\{\overline{U_{tot}(\mathbf{r})} | \boldsymbol{\Theta}(t)\}$. Before proceeding further, we introduce three lemmas which will be used in the analysis of the performance of DLSA.

**_Lemma 1._** Under any feasible data admission action, power allocation action, routing and scheduling action, and link selection action that can be implemented at time $t$, we have:

$$\Delta_V(t) \leq B - \mathbb{E}\{\sum_{n,c}[VU_n^c(R_n^c(t)) - Q_n^c(t)R_n^c(t)] | \mathbf{Q}(t)\} \\ - \mathbb{E}\{\sum_n[\sum_{c,b}\mu_{nb}^c(t)\gamma_{nb}(Q_n^c(t) - Q_b^c(t)) - Z_n(t)y_n(t)] | \boldsymbol{\Theta}(t)\} \quad (12)$$

where $y_n(t) = \sum_b p_{nb}\gamma_{nb} - P_{tot}$, and $B$ is a constant satisfying $B = N^2\left(\frac{3}{2}(d^i)^2 \mu_{\max}^2 + R_{\max}^2\right) + N\frac{1}{2}(P_{\max} + P_{tot})^2$.

**Proof**

First by squaring both sides of (5), and using the fact that $\left([x]^+\right)^2 \leq x^2 \ \forall x \in \mathbb{R}$, we have:

$$[Q_n^c(t+1)]^2 - [Q_n^c(t)]^2 \leq \left[\sum_{b \in \mathcal{N}_i^o} \mu_{nb}^c(t)\right]^2 + \left[\sum_{a \in \mathcal{N}_i^i} \mu_{an}^c(t) + R_n^c(t)\right]^2 \\ - 2Q_n^c(t)[\sum_{b \in \mathcal{N}_i^o} \mu_{nb}^c(t) - \sum_{a \in \mathcal{N}_i^i} \mu_{an}^c(t) - R_n^c(t)] \tag{13}$$

Multiplying both sides by $\frac{1}{2}$, and defining $\hat{B} = \frac{3}{2}(d^i)^2 \mu_{\max}^2 + R_{\max}^2$, we have:

$$\frac{1}{2}([Q_n^c(t+1)]^2 - [Q_n^c(t)]^2) \leq \hat{B} - Q_n^c(t)[\sum_{b \in \mathcal{N}_i^o} \mu_{nb}^c(t) - \sum_{a \in \mathcal{N}_i^i} \mu_{an}^c(t) - R_n^c(t)] \tag{14}$$

Using a similar approach, we get that:

$$\frac{1}{2}([Z_n(t+1)]^2 - [Z_n(t)]^2) \leq \hat{B}' + Z_n(t)[\sum_{b \in \mathcal{N}_n^o} p_{nb}\gamma_{nb} - P_{tot}], \text{ where } \hat{B}' = \frac{1}{2}(P_{\max} + P_{tot})^2 \tag{15}$$

Now by summing (14) over all $(n,c)$ and (15) over all $n$, and by defining $B = N^2\left(\frac{3}{2}(d^i)^2 \mu_{\max}^2 + R_{\max}^2\right) + N\frac{1}{2}(P_{\max} + P_{tot})^2$, we have:

$$L(t+1) - L(t) \leq B + \sum_{n,c} Q_n^c(t)R_n^c(t) - \sum_n [\sum_{c,b} \mu_{nb}^c(t)\gamma_{nb}(Q_n^c(t) - Q_b^c(t)) - Z_n(t)y_n(t)] \tag{16}$$

Taking expectations on both sides over the random channel states and the randomness over actions conditioning on $\Theta(t)$, subtracting from both sides the term $\mathbb{E}\{U_{tot}(\bar{\mathbf{r}})|\Theta(t)\}$, and rearranging the terms, we see that Lemma 1 follows. □

**_Lemma 2._** We use $r^*$ to denote the optimal solution of the problem (8). The optimal network utility $U_{tot}(r^*)$ satisfies: $VU_{tot}(r^*) \leq \phi^*$, where $\phi^*$ is obtained over the following optimization problem using deterministic strategies and stochastic strategies based only on the observed $\mathbf{S}(t)$: allocate constant admission rates $r^{nc}$ every timeslot; when $\mathbf{S}(t) = s_i$, choose a power matrix $\mathbf{P}_m^{s_i}(t)$ and link connection matrix $\mathbf{\Gamma}_n^{s_i}(t)$ with probability $\rho_m^{s_i}$ and $\phi_n^{s_i}$, subject to (4), (5) and other constraints mentioned in Section 2, allocate service rate $\mu_{ij}^c(s_i, \mathbf{\Gamma}_n^{s_i}(t), \mathbf{P}_m^{s_i}(t))$ to node $n$ to satisfy:

$$\max \quad \phi = VU_{tot}(r^*)$$
$$\text{s.t.} \quad r^{nc} + \mathbb{E}\{\sum_{m,n} \rho_m^{s_i}\phi_n^{s_i} \sum_{a \in \mathcal{N}_i^i} \mu_{an}^c(t)\} \leq \mathbb{E}\{\sum_{m,n} \rho_m^{s_i}\phi_n^{s_i} \sum_{b \in \mathcal{N}_i^o} \mu_{nb}^c(t)\} \tag{17}$$
$$\mathbb{E}\{\sum_{m,n} \rho_m^{s_i}\phi_n^{s_i} \sum_{b \in \mathcal{N}_i^o} p_{nb}\gamma_{nb}\} \leq P_{tot}$$

Here the expectation is taken over the random channel state $s_i$.

**Proof**

Proof can be found in Chap 4 in [11] and [15], and is omitted here for brevity. □

**_Lemma 3._** Suppose there are constants $B \geq 0, V \geq 0$ and $y_0$ such that for all timeslots $t$, and all possible values of $\Theta(t)$, we have $\Delta_V(t) \leq B - Vy_0$, then all queues are mean rate stable.

**Proof**

This lemma is a specific form of Theorem 4.2 in [11]. Proof can be found in chap 4.1 in [11] and is omitted here for brevity. □

By using Lemma 1-3, we have the following theorem, which states the network is stable under DLSA and utility function is arbitrarily close to the optimal value.

*Theorem 1*

Under DLSA, the network is mean rate stable. And we have:

$$\sum_{n,c} U_n^c(\limsup_{t\to\infty} \frac{1}{t}\sum_{\tau=0}^{t-1} \mathbb{E}\{R_n^c(t)\}) \geq U_{\text{tot}}(r^*) - \frac{B}{V},$$

which mean the gap between utility function achieve by DLSA and the optimal value under any other is not more than $B/V$.

**Proof**

Define the RHS of (12) without expectation in Lemma 1 as

$$D(t) = \sum_{n,c}[VU_n^c(R_n^c(t)) - Q_n^c(t)R_n^c(t)] + \sum_n[\sum_{c,b}\mu_{nb}^c(t)\gamma_{nb}(Q_n^c(t) - Q_b^c(t)) - Z_n(t)y_n(t)] \quad (18)$$

Denote $D^{\text{DLSA}}(t)$ and $D'(t)$ as $D(t)$ achieved by DLSA and any other scheduling algorithm. From the description of DLSA, we can see that DLSA maximizes $D(t)$. Hence it follows that $D^{\text{DLSA}}(t) \geq D'(t)$.

Using the algorithm mentioned in Lemma 2 and $VU_{\text{tot}}(r^*) \leq \phi^*$, taking expectations on both sides, we have

$$\mathbb{E}\{D^*(t)\} = \mathbb{E}\{\sum_{n,c}[VU_n^c(R_n^c)) - Q_n^c(t)R_n^c(t)] + \sum_n[\sum_{c,b}\mu_{nb}^c(t)\gamma_{nb}(Q_n^c(t) - Q_b^c(t)) - Z_n(t)y_n(t)]\}$$
$$\geq \phi^* \geq VU_{\text{tot}}(r^*) \quad (19)$$

Hence, for DLSA, we have $\Delta_V(t) \leq B - \mathbb{E}\{D^{\text{DSLA}}(t)\} \leq B - VU_{\text{tot}}(r^*)$. This is in the form of Lemma 3, so all queue are mean rate stable. By the virtual queue technique, time-average power constraint is satisfied.

Summing the above over $t = 0,1,\ldots,T-1$, we have:

$$\mathbb{E}\{L(T) - L(0)\} - V\sum_{t=0}^{T-1}\mathbb{E}\{U_{\text{tot}}(\bar{\mathbf{r}}) | \Theta(t)\} \leq TB - TVU_{\text{tot}}(r^*).$$

Rearranging the terms, using the facts that $L(T) \geq 0$ and $L(0)=0$, dividing both sides by $TV$, and taking the liminf as $t \to \infty$, we have

$$\limsup_{t\to\infty} \frac{1}{t}\sum_{\tau=0}^{t-1}\mathbb{E}\{\sum_{n,c}U_n^c(R_n^c(t))\} \geq U_{\text{tot}}(r^*) - \frac{B}{V}.$$

Using Jensen's inequality, we see that:

$$\sum_{n,c}U_n^c(\limsup_{t\to\infty}\frac{1}{t}\sum_{\tau=0}^{t-1}\mathbb{E}\{R_n^c(t)\}) \geq U_{\text{tot}}(r^*) - \frac{B}{V}.$$

This completes the proof of Theorem 1. □

# 5. Simulation

To verify the effectiveness of our algorithm, we evaluate our algorithm using MATLAB simulator tool.

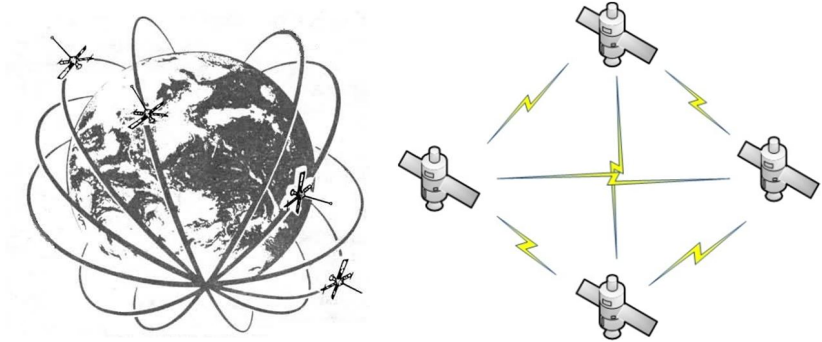

Fig.1 A Satellite Network of 4 Nodes

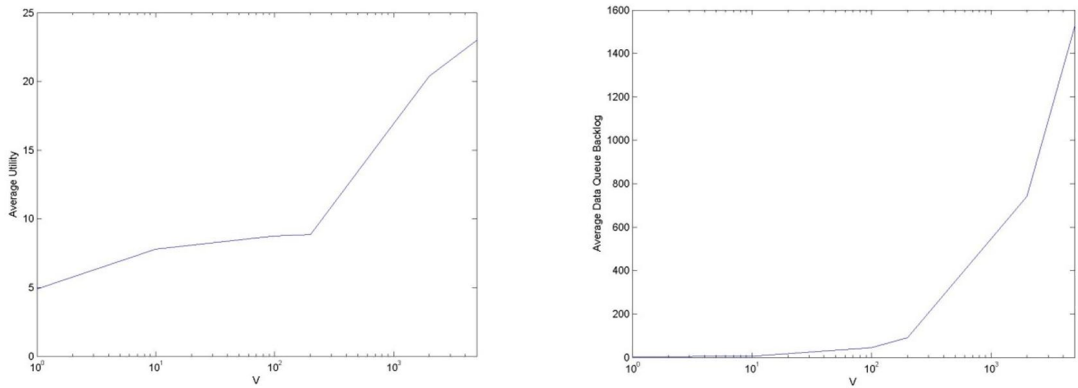

Fig.2 Average Queue Backlog and Utility Function

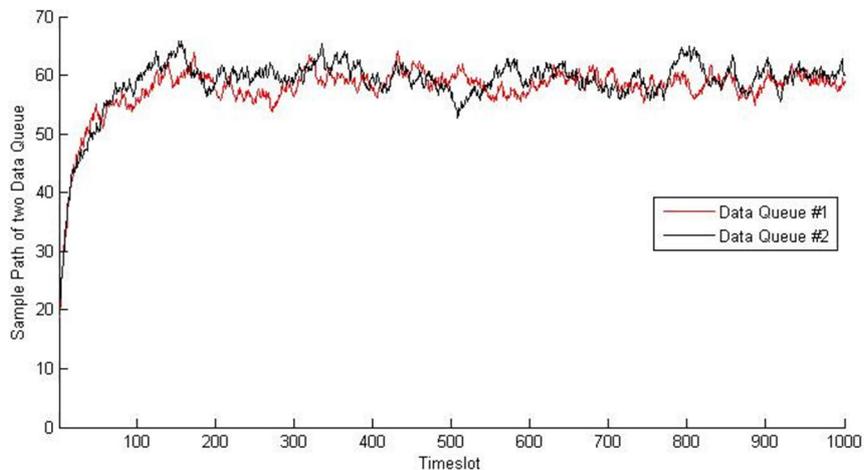

Fig.3 Sample Path of 2 Queues

We consider network consisting of 4 nodes as shown in Fig. 1. Each node communicates with other nodes, which means $\mathcal{N}_n^o = \mathcal{N}_n^i = \{N_j \mid j \neq n\}\ \forall n$. The link state of each communication link is i.i.d. every timeslot and can be either of {"G(Good)", "B(Bad)", "C(Common)", "U(Unreachable)"} with equal probabilities. For link $(i,j)$, define link state factor $\alpha_{ij}(t) = 3, 1, 2, 0$ when $s_{ij}(t) = $ "G", "B", "C", "U" respectively. We assume $R_{\max} = 6$ and the utility functions are given by: $U_n^c(r^{nc}) = \ln(1 + r^{nc})$. We assume $P_i^{\mathrm{av}} = 4, \forall i$, $P_{\max} = 6$ and $d^o = d^i = 2$. For simplicity, we also assume that all the links do not interfere with each other. Rate-power function is assumed in the form $\mu_{ij}(t) = \ln(1 + \alpha_{ij}(t) p_{ij}(t) \gamma_{ij}(t))$.

We simulate as $V$ takes different value in the set $\{1, 10, 100, 200, 1000, 5000\}$. Each simulation is run for $10^6$ slots. The simulation results of average queue backlog and average utility function v.s. different value of $V$ in semi-log scale are plotted in Fig. 2. We see that both the total network utility and average data queue size grow as $V$ increases.

Fig. 3 also shows two sample-path data queue processes under $V=100$ (for the first 1000 timeslots only). We can tell from Fig.3 that the system is stable because the queues are bounded by a finite value.

## 5. Conclusion

In this paper, we proposed a joint scheduling allocation scheme named Degree-Limited Scheduling Algorithm (DLSA), which consists of data admission, power allocation, link selection, and routing. DLSA aims at achieving the maximal value of utility function while stabilizing the network without requiring any stochastic knowledge of link states. After discussion about the implementation, we analyze the performance of DSLA, showing that the network is mean rate stable and utility function is arbitrarily close to the maximal value on the order of O (1/V). Simulation results verify DLSA and our analysis. In future work, we will take effort to reduce the queue backlog at a minor cost of reduction of utility, in order to reduce the delay, which will not only benefit the users but also reduces the storage pressure of this satellite system.

## Acknowledgement

The work described in this paper was supported by the National Natural Science Foundation of China (No. 60902049 and No. 61272518) and Important National Science & Technology Specific Projects (No. 2011ZX03001-004-02).